\documentclass[aps,pre,showpacs,floatfix]{revtex4}
\usepackage{epsf}
\usepackage{subfigure}
\usepackage{amsmath}
\usepackage{amssymb}
\usepackage{graphicx}
\usepackage{epstopdf}
\newcommand{\be}{\begin{equation}}
\newcommand{\ee}{\end{equation}}
\newcommand{\bea}{\begin{eqnarray}}
\newcommand{\eea}{\end{eqnarray}}

\begin{document}

Europhys. Lett. ${\bf 81} \ (2008), \ 28004$.
\vspace{1.5cm}

\title{\bf Dynamical randomness, information, and Landauer's principle}
\author{D. Andrieux and P. Gaspard}
\affiliation{Center for Nonlinear Phenomena and Complex Systems,\\
Universit\'e Libre de Bruxelles, Code Postal 231, Campus Plaine,
B-1050 Brussels, Belgium}
\begin{abstract}
New concepts from nonequilibrium thermodynamics are used to show that 
Landauer's principle can be understood in terms of time asymmetry in 
the dynamical randomness generated by the physical process
of the 
erasure of digital information. In this way, Landauer's principle is 
generalized, showing that the dissipation associated with the erasure 
of a sequence of bits produces entropy at the rate $k_{{\rm B}}I$ per 
erased bit, where $I$ is Shannon's information per bit.
\end{abstract}
\maketitle


Landauer's principle says that the minimal dissipation accompanying
the erasure of one bit of information produces an entropy equal to
$k_{\rm B} \ln 2$ where $k_{\rm B}$ is Boltzmann's constant
\cite{L61}. This important result is based on the observation that
the processing of digital information is a physical process among
others and should thus obey the laws of thermodynamics.  Landauer's
result has therefore supported the idea that information processing
devices working at temperature $T$ should dissipate at least $k_{\rm
B} T \ln 2$ of energy during an elementary act of information
\cite{B62,vN66}.
Later, Bennett's work clarified the point that
Landauer's dissipation is the feature of logically irreversible
operations on data, i.e., operations whose inverse is not unique,
such as data erasure \cite{B73}.  Moreover, Bennett used Landauer's
result to resolve Maxwell's demon paradox \cite{B82}. Since then,
Landauer's principle has been explicitly verified in specific cases,
for instance, in bistable potentials with white noise \cite{S95}, for
a bit in contact with a thermal reservoir  \cite{P00}, or from coarse
graining in phase space \cite{KPV07}.

Landauer's result has been
verified using case studies and it is only recently that
nonequilibrium statistical mechanics has been sufficiently advanced
in order to reach its understanding in a general framework. Indeed,
it has been recently established that the second law of
thermodynamics finds its origin in the time asymmetry of the property
of dynamical randomness, i.e., the temporal disorder that a physical
process develops during its time evolution \cite{G04,G05,AGCGJP07,G07}.
Dynamical randomness is characterized by the decay rate of the
probabilities of the typical paths or histories followed by the
fluctuating process.  The time asymmetry of the process can be tested
by comparing the decay rate of the path probabilities with the one of
the corresponding time-reversed process.  These decay rates are the
so-called entropy per unit time
\bea
h = \lim_{n \rightarrow \infty} - \frac{1}{n}
\sum_{\omega_1\omega_2\cdots\omega_n}
\mu(\omega_1\omega_2\cdots\omega_n) \ln \mu(\omega_1\omega_2\cdots
\omega_n)
\label{h}
\eea
and time-reversed entropy per unit time
\bea
h^{\rm R} = \lim_{n \rightarrow \infty} - \frac{1}{n}
\sum_{\omega_1\omega_2\cdots\omega_n}
\mu(\omega_1\omega_2\cdots\omega_n) \ln
\bar{\mu}(\omega_n\cdots\omega_2\omega_1)
\label{hR}
\eea
where $\omega_1\omega_2\cdots\omega_n$ denotes a path. $\mu$ is
the invariant probability measure of the equilibrium or
nonequilibrium process, whereas $\bar{\mu}$ is the invariant measure 
of the backward process with
reversed nonequilibrium drives 
\cite{G04,G05,AGCGJP07}.
The paths are observed every time interval
$\tau$ (considered as the unit time) and sampled into states
$\omega_j$ (with $j=\cdots-2,-1,0,1,2,\cdots$). These dynamical
entropies may be considered as rates of production of information
generated by the fluctuations of the process during its time
evolution.
In this sense, they characterize the dynamical randomness
of the process. It has been established \cite{G04,G05,AGCGJP07} that 
the difference between
these quantities gives the well-known thermodynamic entropy production
according to
\bea
\Delta_{\rm i} S = k_{\rm B} (h^{\rm R} -h)
\label{dis}
\eea
where $\Delta_{\rm i} S$ is the entropy produced during the unit time
$\tau$. The difference between the two entropies per unit 
time is the relative entropy between the forward and backward paths,
which is known to be non-negative and to give the thermodynamic 
entropy production 
for instance in Markovian stochastic processes 
\cite{G04,G05,AGCGJP07}.
This result has been verified in experiments on driven
Brownian motion and electric noise \cite{AGCGJP07}.
Equation (\ref{dis}) explicitly shows that the thermodynamic entropy
production comes from the time asymmetry in the more microscopic
property of dynamical randomness. The
non-negativity of the entropy production leads to the principle of
temporal ordering according to which, in nonequilibrium steady
states, the typical paths are more ordered in time than their
corresponding time reversals \cite{G07}.

The remarkable result we will here report is that the
aforementioned formula (\ref{dis}) allows us to relate the
thermodynamic entropy production to the information as it is
physically recorded {\it in space} inside some information processing
device, generalizing in this way Landauer's principle.  We here
consider the process of erasure of {\it statistically correlated}
random bits, that is a sequence of bits
$\sigma_1\sigma_2\cdots\sigma_m\cdots$ (with $\sigma_i=0$ or $1$).
We 
assume that this sequence is initially recorded 
on some spatially 
extended support such as a
recording tape inside the device. 
The recorded data are described 
by some probability
distribution $p(\sigma_1\sigma_2\cdots\sigma_m)$ 
giving the 
occurrence frequencies of the 
sequences
$\sigma_1\sigma_2\cdots\sigma_m$ in the memory of the 
device.
This probability distribution is general with possible
spatial correlations among the bits $\sigma_i$.
The information 
contained in the
sequence is characterized by the quantity
\bea
I = \lim_{m \rightarrow \infty} - \frac{1}{m}
\sum_{\sigma_1\sigma_2\cdots\sigma_m}
p(\sigma_1\sigma_2\cdots\sigma_m) \ln
p(\sigma_1\sigma_2\cdots\sigma_m)
\eea
This is an entropy per bit of information in the sense of
Shannon, which we call the information $I$ per bit of the sequence.
In the case the bits are randomly distributed with equal probability
and independently of each other, the information per bit is equal to
$I=\ln 2$.  In general, we have the inequality $I \leq \ln 2$.

\begin{figure}[h]
\centerline{\scalebox{0.6}{\includegraphics{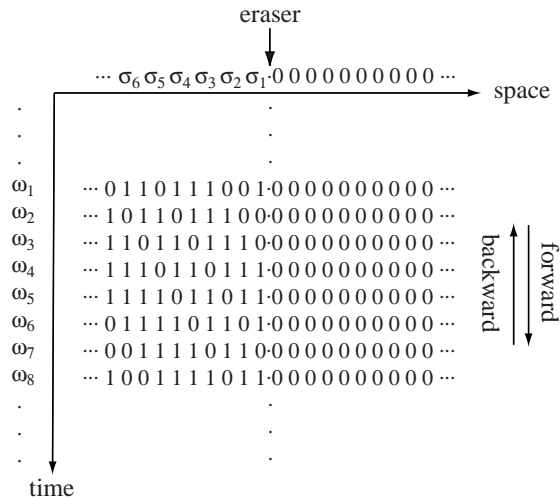}}}
\caption{Space-time plot of the physical process of erasure of a
sequence of bits $\sigma_1\sigma_2\cdots\sigma_m\cdots$ of
information.  The bits are distributed along the
space axis on the
recording tape of the information processing device. The eraser is
located
somewhere along the recording tape and transforms each bit
into a zero
in this illustrative example. At every instant of
time, the state of the system is given by the current sequence of
bits: $\omega_j=\cdots\sigma_{j+2}\sigma_{j+1}\sigma_{j}\cdot
00000\cdots$ where the dot denotes the location of the eraser.}
\label{fig1}
\end{figure}

The thermodynamic entropy produced during the process of erasure of the
aforementioned sequence can be obtained using the formula (\ref{dis})
according to the following reasoning.  We suppose that one bit is
erased every unit time $\tau$.
Let us associate a state $\omega_i$ with the sequence of 
bits $(\cdots\sigma_m\cdots\sigma_{i+1}\sigma_i\cdot 000\cdots)$, as 
done in Fig. \ref{fig1}.
When viewed forward in time, the erasure process transforms the state 
$\omega_i$ into $\omega_{i+1}$.
Therefore, the process of erasure does not generate dynamical 
randomness since the
outcome is unique every time a bit is erased.  Accordingly, the
dynamical entropy per unit time (\ref{h}) vanishes, $h=0$, since the 
probability measure $\mu$
takes the unit value for the unique path 
followed during erasure and vanishes for all the other paths. 
On the 
other hand, the backward process corresponds to the generation of a 
sequence of bits. This
reversed process is not unique and generates dynamical randomness at 
the rate $h^{\rm
R}=I$ given by the information contained in the sequence of bits (see
Fig. \ref{fig1}).
Indeed, the sequence of bits $(\sigma_1\sigma_2\cdots\sigma_m)$ now 
appears 
at random with the probability 
$p(\sigma_1\sigma_2\cdots\sigma_m)$ 
and the probability measure 
$\bar{\mu}$ 
of this reversed process is distributed among several 
possible time-reversed paths.
This observation can be understood by the following example. Let us 
consider a particle in a bistable potential, where the left and right 
wells correspond to the bits $0$ and $1$. We can slowly deform the 
potential in order to force the particle to end in the left well, 
which is equivalent to the erasure of the initial bit. However, since 
any such deformation must pass through a potential with a single 
minimum, undoing the latter transformation will result in the 
particle being in the left or right well with equal probability. This 
is the analog of our backward process, where the sequences of bits 
are now generated according to their respective probabilities. In 
this way, we can understand how dissipation is closely related to 
logical irreversibility \cite{L61}. Finally, we infer from Eq. 
(\ref{dis}) that the
thermodynamic entropy production of the erasure is given by
\bea
\Delta_{\rm i} S = k_{\rm B}( h^{\rm R} -h) = k_{\rm B} I \ \text{per bit} \ .
\label{gen.land.}
\eea
Landauer's principle is recovered in the particular case of
statistically independent random bits of equal probability for which
$\Delta_{\rm i} S = k_{\rm B}I = k_{\rm B}\ln 2$ \cite{G04}.
This
shows that Landauer's principle can be understood from concepts
uniquely based on dynamical randomness. Therefore, it is completely
model independent and $k_{{\rm B}} I$ is the {\it minimal} dissipation
one can achieve for correlated random bits. Another way to obtain
this result is to note that there exist universal coding schemes that
will asymptotically compress any ergodic sequence of length $m$ to
its maximal possible value $m I_2$ (the subscript $2$ indicates that
the information is calculated with logarithms in base $2$) \cite{CT91}. Once
compressed, the sequence is composed of uncorrelated random bits with
relative probability one half (otherwise it would be possible to
further compress the sequence, leading to a compression smaller than
$I_2$ in contradiction with Shannon's bound on data compression). The
usual Landauer principle can then be applied to this compressed
sequence which leads to a dissipation of $k_{\rm B} I_2 \ln 2 =
k_{\rm B} I$ per bit.

To avoid this computational step in the case where the exact probability distribution $p$ of 
the bits is unknown, we can still obtain the minimal entropy 
production (\ref{gen.land.}) as follows.
The extra cost of compressing data using an {\it a priori} 
distribution $q$ instead of the correct distribution $p$ is the 
relative entropy $D(p\|q)$,
given by \cite{CT91}
\bea
D(p\|q)=\lim_{m \rightarrow \infty} \frac{1}{m}
\sum_{\sigma_1\sigma_2\cdots\sigma_m}
p(\sigma_1\sigma_2\cdots\sigma_m) \ln \frac{
p(\sigma_1\sigma_2\cdots\sigma_m)}{q(\sigma_1\sigma_2\cdots\sigma_m)}
\eea
It has been proven in Ref. \cite{X98} that there exists a stationary 
ergodic process $\bar{q}$ with the property to have a vanishing 
relative entropy, $D(p\|\bar {q})=0$, with respect to any other 
stationary ergodic process with probability distribution $p$. 
Therefore, the erasure process can be done assuming this particular 
distribution, which results in no extra cost with respect to the 
minimal entropy production (\ref{gen.land.}).

As long as Landauer's principle is equivalent to the second law of
thermodynamics \cite{B03}, the agreement between the argument based
on dynamical randomness and the universal compression procedure shows
that Eq. (\ref{dis}) is the appropriate measure of dissipation for
general ergodic stationary stochastic
processes.

We notice that the information $I$ is positive if the probabilities
$p(\sigma_1\sigma_2\cdots\sigma_m)$ characterizing the sequence of
bits decay exponentially.  Nevertheless, information can be
sporadically distributed along the sequence of bits, in which case
the probabilities $p(\sigma_1\sigma_2\cdots\sigma_m)$ decay as
stretched exponentials so that the information per bit vanishes $I\to
0$ in the long-sequence limit $m\to\infty$. The fact is that such
sporadic sequences are not uncommon in complex systems
\cite{GW88,EN91,EN92,NG94}.  Accordingly, the above considerations
lead to the interesting result that dissipation can be arbitrarily
small during the erasure of sequences of sporadically distributed
information.

In conclusion, we have here obtained a generalization
of Landauer's principle on the basis of recent advances in
nonequilibrium statistical mechanics showing that the second law of
thermodynamics can be understand in terms of time asymmetry in the
property of dynamical randomness \cite{G04,G05,AGCGJP07,G07}.  These
advances allow us to relate the thermodynamic entropy production
during the erasure of a sequence of bits to the information contained
in this sequence, taking into account the possible statistical
correlations among the bits. This result clearly demonstrates that
the second law of thermodynamics has fundamental implications on the
way information is processed in physical systems.\\

{\bf Acknowledgments.} This research is financially supported by the
F.R.S.-FNRS Belgium
and the ``Communaut\'e fran\c caise de Belgique'' (contract
``Actions de Recherche Concert\'ees'' No. 04/09-312).

\end{document}